\documentstyle[twocolumn,aps]{revtex}
\input epsf
\begin{document}
\twocolumn[\hsize\textwidth\columnwidth\hsize\csname@twocolumnfalse\endcsname

\title{High-frequency Cross-correlation in a Set of Stocks}
\author{Giovanni Bonanno$^{1}$, Fabrizio Lillo$^{1}$
and Rosario N. Mantegna$^{1,2}$} 
\address{
$^1$ Istituto Nazionale per la Fisica della Materia, 
Unit\`a di Palermo, Facolt\`a di Ingegneria, Universit\`a di Palermo,\\
Viale delle Scienze, I-90128 Palermo, Italy\\
 and \\
$^2$ Dipartimento di Fisica e Tecnologie Relative, 
Universit\`a di Palermo, Viale delle Scienze, I-90128 Palermo, Italy}

\maketitle 

\begin{abstract} 

The high-frequency cross-correlation existing between pairs of stocks
traded in a financial market are investigated in a set of 100 stocks 
traded in US equity markets. A 
hierarchical organization of the investigated stocks is obtained by 
determining a metric distance between stocks and by investigating 
the properties of the subdominant ultrametric associated with it.
A clear modification of the hierarchical organization 
of the set of stocks investigated is detected when 
the time horizon used to determine stock returns is changed.
The hierarchical location of stocks of the energy sector is
investigated as a function of the time horizon.

\end{abstract}

\vskip2pc]

\narrowtext
\section{INTRODUCTION}

The presence of a high degree of cross-correlation between
the synchronous time evolution of a set of stocks 
is a well known empirical fact observed in financial markets
\cite{Markowitz59,Elton95,Campbell97}. The typical time 
horizon used to determine the cross-covariance or 
the correlation coefficient of a pair of time series of the 
price of a financial asset is one trading day. 
For this time horizon correlation coefficients as high as 0.7 
can be observed for some pair of stocks belonging to the same 
economic sector. 

The process of clustering a set of economic entities can 
improve economic forecasting and modeling of composed financial
entities such as, for example, stock portfolios. The most common
method of forming homogeneous groups is the principal 
components analysis of the correlation matrix of the raw data
\cite{Elton71}. Correlation based clustering of synchronous
financial data sampled at a fixed time horizon 
has also been performed
to obtain a taxonomy of financial data. 
This has been done by using as a similarity measure the 
correlation coefficient between two financial time series 
\cite{Panton76}.

Recently \cite{Mantegna99}, it has been proposed to detect
the economic information present in a correlation coefficient 
matrix with a filtering procedure based on the estimation
of the subdominant ultrametric \cite{Rammal86} associated with 
a metric distance obtained starting from the correlation coefficient
matrix of a set of $n$ stocks. A metric distance can be obtained 
starting from the correlation coefficient as indicated in Ref. 
\cite{Gower66}.  Having obtained a metric distance one can
obtain the minimum 
spanning tree (MST) and a hierarchical tree 
associated to each correlation coefficient matrix
by using the nearest neighbor single linkage cluster analysis
\cite{Mardia79}. 
In other words, geometrical (throughout the MST) and 
taxonomic (throughout the hierarchical tree)
aspects of the correlation present between the
stock pairs of a set of stocks can be  sorted out 
using the information contained in
the correlation coefficient matrix. To the best of our knowledge
this correlation based clustering method was used for the 
first time to investigate financial data in Ref. \cite{Mantegna99}.

It is known since 1979 that the degree of intrasector 
cross-correlation diminishes by diminishing the time 
horizon used to compute 
stock returns \cite{Epps79}. This phenomenon is sometime 
addressed as ``Epps effect". The existence of this phenomenon 
motivates us to investigate the nature and the properties 
of the hierarchical organization obtained starting from 
the correlation coefficient matrix as a function of the time 
horizon used to obtain it.
 
In the present study, we use the high-frequency data of the transactions
occurring in the US equity markets which are recorded in the 
{\it Trade and Quote}
database of the New York Stock Exchange. By using this database we
are able to investigate comovements of a set of highly capitalized 
stocks for daily and intra daily time horizons.

The decrease of the cross-correlation between the stock returns observed
for diminishing time horizons
progressively changes the nature of the hierarchical structure
associated to each different time horizon.
The structure of the considered set of stocks 
changes moving from a complex 
organization to a progressively elementary one when the time horizon 
of price changes varies from $d=6$ h and $30$ min to $d/20=19$ 
min and 30 sec, where $d$
is the daily time horizon at the New York Stock Exchange.

\section{Hierarchical structure in a portfolio of stocks}

Correlation matrices of price changes 
have been recently investigated by physicists within
the framework of random matrix theory \cite{Laloux99,Plerou99}.
Here we use the method proposed in Ref. \cite{Mantegna99}
to filter the economic information stored 
in the correlation coefficient matrix in a way which is 
straightforward and independent of any external threshold. 
Specifically, the filtering is done (i) by starting from the 
synchronous correlation coefficient of the difference of 
logarithm of stock price computed at a selected time horizon, 
(ii) by calculating a metric distance 
between all the pair of stocks and (iii) by selecting the 
subdominant ultrametric distance associated to the 
considered metric distance.
 
The correlation coefficient is defined as 
\begin{equation}
\rho_{ij} (\Delta t) \equiv \frac{<Y_i Y_j>-<Y_i><Y_j>}
{\sqrt{(<Y_i^2>-<Y_i>^2)(<Y_j^2>-<Y_j>^2)}}
\end{equation}
where $i$ and $j$ are  numerical labels of the stocks, 
$Y_i=\ln S_i(t)-\ln S_i(t-\Delta t)$, $S_i(t)$ is the value of 
the stock price $i$ at the trading time $t$ and $\Delta t$ is the
time horizon. 
For each value of $\Delta t$, the correlation coefficient 
for logarithm price differences (which almost coincides with 
stock returns) is 
computed between all the possible pairs of stocks present in 
the considered portfolio. 
The empirical statistical average, indicated in this paper 
with the symbol $<.>$,
is here a temporal average always
performed over the investigated time period. 

By definition, $\rho_{ij} (\Delta t)$ can vary from -1 (completely 
anti-correlated pair of stocks) to 1 (completely correlated 
pair of stocks).
When $\rho_{ij} (\Delta t)=0$ the two stocks are uncorrelated. 
The matrix of correlation coefficient is a symmetric matrix with 
$\rho_{ii}(\Delta t)=1$ in the main diagonal. 
Hence for each value of $\Delta t$, $n~(n-1)/2=4950$ correlation 
coefficients characterize 
each correlation coefficient matrix completely.

A metric distance between pair of stocks can be rigorously 
determined \cite{Gower66} by defining
\begin{equation}
d_{i,j} (\Delta t)=\sqrt{2(1-\rho_{ij}(\Delta t))} .
\end{equation}
With this choice $d_{i,j}(\Delta t)$ fulfills the three axioms of a metric --
(i) $d_{i,j}(\Delta t)=0$ if and only if $i=j$; 
(ii) $d_{i,j}(\Delta t)=d_{j,i} (\Delta t)$ and (iii)
$d_{i,j} (\Delta t) \le d_{i,k} (\Delta t) +d_{k,j} (\Delta t)$. 
The distance matrix ${\bf{D}} (\Delta t)$ is 
then used to determine the MST connecting the $n$ stocks. 

The MST, a theoretical concept of graph theory 
\cite{West96}, is a graph with $n-1$ links which selects 
the most relevant connections of 
each element of the set. The MST allows to obtain, 
in a direct and essentially unique way, the subdominant 
ultrametric distance matrix ${\bf{D}}^<(\Delta t)$ 
and the hierarchical organization of the elements 
(stocks in our case) of the investigated data set.

The subdominant ultrametric distance between $i$ and $j$
objects, i.e. the element $d^<_{i,j}$ of the ${\bf{D}}^<(\Delta t)$  
matrix, is the maximum value 
of the metric distance $d_{k,l}$ detected by moving 
in single steps from $i$ to $j$ through the path connecting 
$i$ and $j$ in the MST. 
The method of constructing a MST linking a set of $n$ objects 
is direct and it is known in multivariate analysis as the
nearest neighbor single linkage cluster analysis \cite{Mardia79}. 
A pedagogical exposition of the determination of the
MST in the contest of financial time series is provided in Ref. 
\cite{MS2000}.

Subdominant ultrametric space \cite{Rammal86} has been 
fruitfully used in the description of frustrated complex systems. 
The archetype of this kind of systems is a spin glass \cite{Mezard87}. 
In Ref. \cite{Mantegna99}  the `ansatz' that 
the subdominant ultrametric space may reveal part of the 
economic information stored in the time series of the 
investigated set was proposed. 

\section{Empirical results}

We investigate a set of 100 highly capitalized stocks traded in 
the major US equity markets during the period January 1995 - 
December 1998. At that time, most of them were  used to compute the
Standard and Poor's 100 index. The prices are transaction prices 
stored in the {\it Trade and Quote} database of the 
New York Stock Exchange. The amount of information processed 
consists of about 100 millions of transactions.

The time horizons investigated in the present study 
are $\Delta t=d=6$ h and $30$ min (a trading day time interval), 
$\Delta t=d/2=3$ h and $15$ min, $\Delta t=d/5=1$ h and $18$ min,
$\Delta t=d/10=39$ min and $\Delta t=d/20=19$ min and $30$ sec.
With this choice, in addition to the daily time horizon,
we also investigate intra-day time horizons which are
progressively shorter.

The daily mean number of transactions for the 100 selected stocks
is ranging from 11944.3 transactions of Intel Corp. (INTC)
to the 121.48 transactions of Mallinckrodt Inc. New (MKG).
In the following we will use the tick symbols to label the 
various stocks composing our set. Information about the 
company indicated by each tick symbol can be easily find
in several financial web pages such as, for example, 
http://www.quicken.com .
To assess the typical amount of the spreading around 
the mean value observed for the daily number of transactions 
we give the minimum and the maximum daily number of 
transactions occurring for the two companies
considered above during the investigated period. 
Specifically the minimum number of daily transactions 
was 1717 for INTC and 22 for MKG whereas the maximum was 
95437 for INTC and 388 for MKG.
By considering that the minimum number of daily transactions 
is above 17 for the 100 stocks considered in the present study,
we decide to investigate time horizons as short as $d/20$ 
to statistically ensure that at least  1 transaction
occurs during the time horizon $\Delta t$.

In Fig. 1 we show the hierarchical tree
obtained for a $\Delta t=6$ h and $30$ min time horizon. 
In the figure each vertical line indicates 
a stock. The sector of each stock is coded with a color 
code provided in the figure caption. Several clusters are 
clearly identified. The most prominent are (i) the cluster 
of energy stocks (the blue cluster), (ii) the cluster
of financial stocks (green), (iii)  technology cluster (red),
(iv) basic materials cluster (violet) and (v) utilities 
cluster (magenta). Hence a collection of one-day time horizon 
time series of returns carries
information about the economic sector of the stocks considered.
Additional information about the form and structure of the
subdominant ultrametric associated with the metric distance
is present in the MST. In Fig. 2 we show the MST 
obtained for the same time horizon as in Fig. 1. 
In the MST some stocks, such as for example, BAC, INTC or AEP are
linked with several stocks belonging to the same sector 
(financial, technology and utility sector respectively in the 
discussed example) whereas others (most notable case
is General Electric Co., GE) are linked to stocks of different 
sectors. In this
second case the stock with several links acts as a 
`reference' stock at different hierarchically structured 
distances.

The `Epps effect' predicts that the intra-sector pair
correlation decreases by decreasing the time horizon $\Delta t$.
In Fig. 3 we show the mean correlation coefficient $<\rho>$
obtained by averaging over the $n(n-1)/2$ off-diagonal 
elements of the
correlation coefficient matrix. In the same figure, we also show 
the minimum $\rho_{min}$ and the maximum $\rho_{max}$ value of 
the correlation coefficient observed in the correlation
matrix as a function of the time horizon $\Delta t$.  
  
As expected the values of $\rho_{max}$, $|\rho_{min}|$ and 
$<\rho>$ are decreasing when 
$\Delta t$ decreases. The most prominent correlation weakening 
is observed for the most correlated pair of stocks (the ones 
having a correlation coefficient closes to the maximum value 
$\rho_{max}$). In fact $\rho_{max}$ decreases from 0.76
to 0.52 when $\Delta t$ changes from 6 h and 30 min to 
19 min and 30 s.

The decrease of the correlation between pairs 
of stocks affects the nature of the hierarchical organization of 
stocks. The clusters observed in figures 1 and 2 progressively
disappear and the arrangement of the hierarchical and minimum
spanning tree moves from a structured and clustered set to
a simpler set. Figures 4 and 5 show the hierarchical tree 
and the minimum
spanning tree observed when the time horizon is set to the 
minimal value investigated ($\Delta t=19$ min and 30 s). The number
of clusters 
is diminished and also the number of stocks belonging
to the remaining ones has decreased. The biggest clusters that are 
recognizable in figures 4 and 5 are the technology cluster
(red lines in Fig. 4 and stocks at the left side of Fig. 5),
the financial cluster (green lines in Fig. 4 and stocks
on the right of Fig. 5) and two energy clusters (blue lines
in Fig. 4 and stocks at the bottom and on the right top
of Fig. 5). The change of structure of the MST
is indeed dramatic if one considers the role of some `reference'
stock such as, in the present case, GE. It acts as
a hierarchical reference for 17 stocks (17 links in the MST)
when $\Delta t=6$ h and 30 min whereas it acts as
a reference for 61 stocks 
when the time horizon is decreased to $\Delta t=19$ min and 30 s.

It is worth pointing out that the change in the structure of the
MST and hierarchical tree is not just a simple 
consequence of the 
`Epps effect'. In fact the changes observed in the structure 
of the MST and of the hierarchical tree suggest that the
intrasector correlation  decreases
faster than intersector correlation between pairs of stocks
of the considered portfolio. To verify this hypothesis we considered 
the six largest groups of stocks belonging to the same industrial
sector and, for each time horizon, we measured both the intrasector
mean value ($<\rho^{(in)}>$) and the
intersector mean value ($<\rho^{(out)}>$) of the correlation 
coefficient. For a set of $n_s$ stocks belonging to a specific sector
the average of $\rho^{(in)}$ is done by considering $n_s(n_s-1)/2$
correlation coefficient whereas the average of $\rho^{(out)}$ involves
$(n-n_s) n_s$ elements of the correlation coefficient matrix.
In Fig. 6 we show the values of $<\rho^{(in)}>-<\rho^{(out)}>$ 
as a function of $\Delta t$ for the technology ($n_s=17$), services
($n_s=13$), basic materials ($n_s=11$), financial ($n_s=10$),
consumer non-cyclical ($n_s=10$) and energy ($n_s=8$) sector.
In all the considered sectors it is evident that the decrease
of the degree of correlation is faster at an intrasector 
level rather than at an intersector
level. The different modification of the degree of correlation
at intersector or intrasector level confirms that a 
hierarchical structuring of a stock portfolio occurs when 
$\Delta t$ increases.

The change of structure of the hierarchical and minimum spanning tree
occurs gradually with the decrease of the time horizon. To support
this statement, in Table I we report the number of links 
of a group of stocks as a function of the time horizon. The listed
stocks are chosen by selecting the stocks having more than one 
link when $\Delta t=6$ h and 30 min. The Table
shows that several stocks diminish the
number of their links in the minimum spanning tree when the
time horizon decreases from $\Delta t=d$ to 
$\Delta t=d/20$. 
Exceptions to this general trend are expected because 
the total number of links of a MST ($n-1$) is a conserved 
quantity.  
The key exception is GE (General Electric Co).
The number of links of GE increases gradually from 17 to 
61. In other words, the organization of the MST 
in several clusters 
observed in Fig. 2  is gradually lost when the time horizon is
decreased. In parallel to the disruption of local clusters 
a star-like organization with GE at the center of the star
progressively emerges
(see Fig. 5).

To assess the statistical robustness of our results 
we perform several tests devoted to -- (i) check the significance 
of the results obtained by comparing them with the ones obtainable
starting from surrogate data, (ii) estimate the role of the outliers in
the correlation based clustering of the considered portfolio and
(iii) consider the stability of the 
results across time.

In Fig. 7 we show the mean length (top panel) and its  
standard deviation (bottom panel) of the link
of the minimum spanning tree as a function of the time 
horizon $\Delta t$. We investigate  real data and  
surrogate Gaussian data having the same correlation coefficient 
matrix as in the real case for the 
shortest time horizon. Real data are shown as black circles 
whereas empty circles indicates the results obtained with
surrogate data. The error bars shown in the case of surrogate 
data are one standard deviation in the considered variable.
This value is obtained by repeating the generation of surrogate 
data 100 times. The investigation of surrogate data 
clearly shows that the decrease of the distance of stock pairs
as a function of the time horizon is a real effect that 
cannot be mimicked by surrogate data. The same conclusion is reached
when the standard deviation of the link length of
the minimum spanning tree is considered. The bottom panel of
Fig. 2 shows that the increase of the standard deviation of $d^<_{i,j}$
observed in real data, which is a simple measure of the 
increasing complexity of the MST and hierarchical tree,
cannot be reproduced by surrogate data.

The estimation of the correlation coefficient is rather 
sensitive to outliers. To check how our results 
are affected by them we repeat our investigation by 
considering only returns which are in absolute value
less than $k \sigma_{\Delta t}$ where k is set to be 1,2,3,5 and 10
and $\sigma_{\Delta t}$ is the standard deviation of all returns
of the 100 stocks observed for the selected time horizon.
In our test the returns which are, in absolute value, 
larger than the chosen cutoff are set to zero in the 
calculation of the correlation coefficient.
We verify that 65 of 99 links are still observed 
when we consider the rather severe cutoff induced 
by setting k=3 and $\Delta t=19$ min and 30 s.
For the same parameters but for a 1 trading day time 
horizon the number of conserved links is 72.
When the cutoff is progressively increased the number
of conserved links increases up to 90 (for $\Delta t=19$
min and 30 s) and 87 (for $\Delta t=6$ h and 30 min).
In summary the results obtained are not due to the outliers but
rather the large majority of links are reflecting 
the presence of a synchronization down to the scale 
of few standard deviation of the ensemble return. 

The stability of our results across different averaging period
is tested by repeating some of the analyses done by dividing 
the averaging time interval in subintervals. We verify that 
a stability of statistical nature is present in our results. 
For example, by investigating the coordination number of GE yearly 
for $\Delta t=19$ min and 30 s, we observe that this number 
is 19 in 1995, 40 in 1996, 52 in 1997 and 48 in 1998 and that 
the stock GE is always the stock with the largest number of links
of the MST. Hence its role of `hierarchical reference' stock
is confirmed for each year of the time period of our database.

\section{A case-study: the stocks of energy sector}

The progressive structuring of the hierarchical tree as
a function of the time horizon is shown in Fig. 8 for the
representative case of energy sector stocks.
In this figure we show the hierarchical trees 
obtained for the 5 investigated time horizons
and we indicate with black lines 
stocks belonging to the energy sector whereas the remaining
lines indicating all the other stocks are left gray.  

In Fig. 8a ($\Delta t=19$ min and 30 s) seven of the eight stocks belonging to
the energy sector are observed in two distinct clusters. The first 
cluster is composed of XON, MOB, CHV and ARC, whereas the second
comprises SLB,HAL and BHI. The last stock (OXY) is not directly
connected to a stock of the energy sector (it is connected to 
the 'reference' stock GE). 
It may be worth pointing out that the stocks SLB, HAL and BHI
are all working in the sub-sector oil services and equipment
of the energy sector. In Fig. 8b ($\Delta t=39$ min) almost
the same behavior as in Fig. 8a is observed. The only difference being
in the increase of the degree of correlation between stocks
(manifested as a decrease of the ultrametric distance
in the hierarchical tree).
Fig. 8c show the first important change. At this time horizon 
($\Delta t=1$ h and 18 min), seven of the eight stocks of the energy sector
form a single cluster which is distinct from the others up to a 
ultrametric distance of $d^<=1.15$. This cluster splits in two 
sub-clusters composed as (i) SLB, HAL and BHI and (ii) XON, MOB, 
CHV and ARC. In Fig. 8d we observe the complete formation
of the energy sector cluster. In fact when $\Delta t=3$ h and 15 min
all the eight energy stocks belonging to the energy sector form
a single cluster. The cluster is maintaining the internal 
structure as discussed before with the last stock (OXY) joining 
the cluster. In Fig. 8e we show the hierarchical 
tree obtained by using a 1 day time horizon ($\Delta t=6$ h and 30 min).
The energy cluster is still there with a new stock linking at
a longer distance. This stock, indicated as a gray line just at the
right of the energy cluster, is the CGP stock. The CGP stock is classified
by Forbes in the utilities sector and its main sub-sector is natural
gas utilities.
 
This case study shows that the nature, size and interrelation
between the stocks of a given economic sector are affected 
by the value of the time horizon used to compute the stock
return. Specifically time is needed before the cluster 
organization between
stocks of the same sectors takes place completely. Our investigation 
suggests that an intra-day dynamics is present when the process
of the rational 
price formation of a given financial asset is going on.
  
\section{Discussion}

Our investigations of the hierarchical structure observed 
in a set of stocks frequently traded in the main US equity markets
show that the degree and nature of cross-correlation between 
pairs of stock returns varies as a function of the time horizon
used to compute them. Specifically intra-day 
time series of returns show a
lowest degree of correlation that manifest itself in the 
partial disruption of the hierarchical structure observed 
for longer (e.g. daily) time horizons. The 
disruption of the correlation structure is more pronounced 
for intrasector correlation than for intersector 
correlation inducing a modification in the correlation 
based clustering of a stock portfolio whose results become
time horizon dependent.

In fact, at shorter time 
horizons a few stocks (in the present case a single stock)
become a hierarchical reference of the large majority of stock.
The stock that takes this
role is the stock of General Electric Co. 
in the investigated set and time period. The determination
of a hierarchical structure for each time horizon allows to follows
the mechanism of cluster formation as a function of the time horizon.
In the investigated set, the minimum time horizon needed
to observe the complete clustering of the eight energy stocks 
is 3 h and 15 min (half trading day). However, this does not implies that 
until that time horizon correlation between returns of energy stocks
are negligible. Indeed, sub-clusters composing
the overall energy cluster are observed for time horizon as short as 
19 min and 30 s. We interpret these fine details of the cluster formation as
an evidence that the process of rational price formation takes
a finite amount of time to occur.

We conclude by discussing the reason for selecting a similarity measure
which obeys the axioms of a metric distance. By selecting as
a similarity measure a metric distance one ensures that the ultrametric
structure of the MST is the subdominant ultrametric structure.
By translating the mathematical language in common language, 
the detected subdominant ultrametric is the ultrametric 
structure which is closer to the original metric structure. 
This implies that the filtering of information performed by 
moving from the metric distance matrix of $n(n-1)/2$ distinct 
elements to the subdominant ultrametric distance matrix
characterized by $n-1$ elements is the only filtering procedure
based on the single linkage cluster analysis that allows the
minimum lost of information. In other words, under the constraint
of reducing the number of elements of the matrix of interest 
from $n(n-1)/2$ to $n-1$, the use of by a similarity measure which is
a distance ensures that the correlation based clustering retains
after filtering the maximal amount of information 
compatible with the procedure. 

\section{Acknowledgements}
The authors thank INFM and MURST for financial support. This work 
is part of the FRA-INFM project 'Volatility in financial markets'. 
G. Bonanno and F. Lillo acknowledge FSE-INFM for their fellowships.
The authors wish to thank Dr. Roberto Ren\'o for a discussion about 
the ``Epps effect''.

\begin{figure}[t]
\epsfxsize=3in
\epsfbox{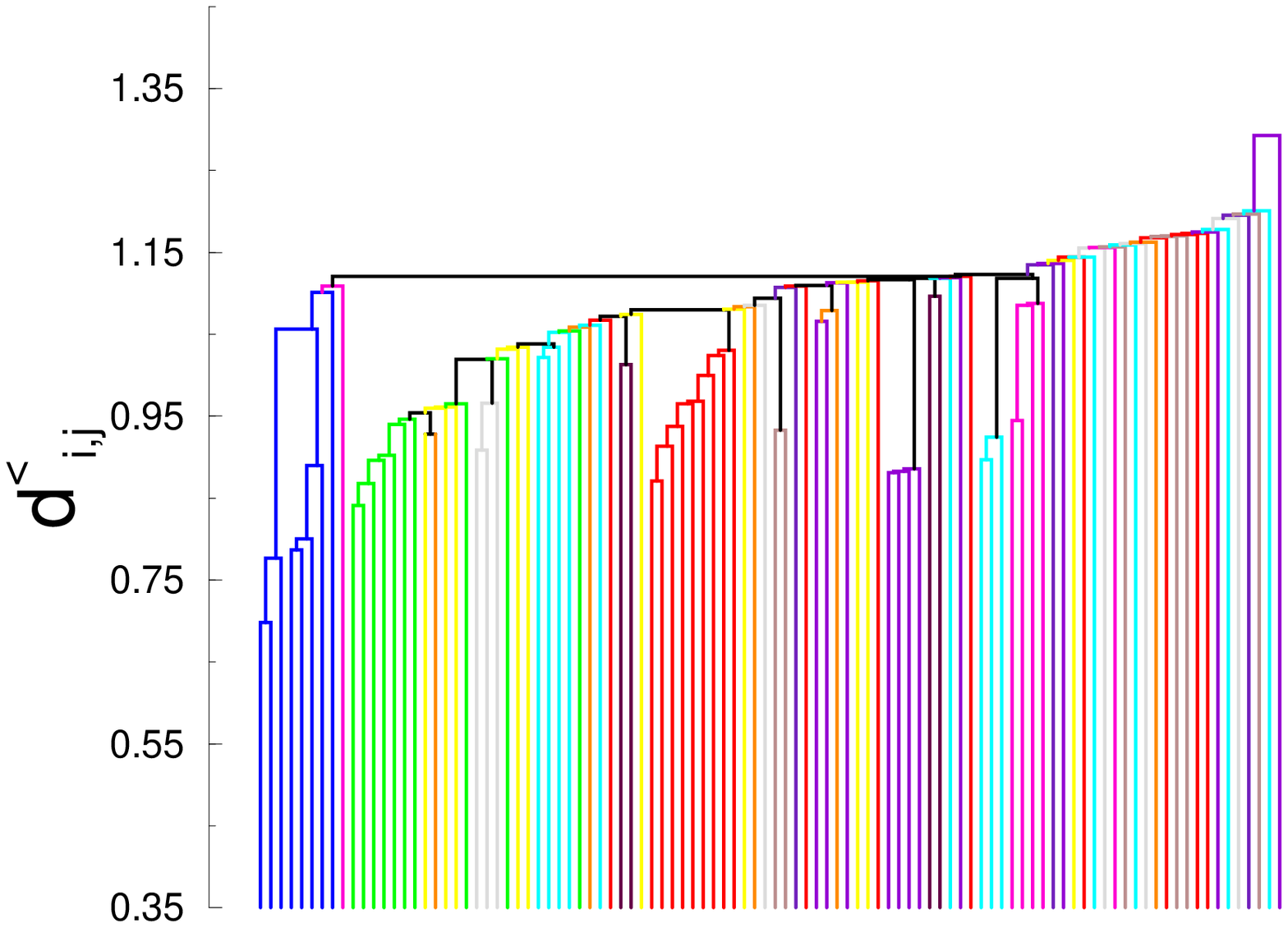}
\vspace{0.3cm}
\caption{Hierarchical tree of the set of 100 stocks traded in the
US equity markets obtained starting from the return time series 
computed with a
$\Delta t=6$ h and 30 min time horizon (1 trading day) during the time period
Jan 1995-Dec 1998. Each stock is indicated by a vertical line and the
color of the line refers to its economic sector as defined by Forbes company.
The presence of several clusters of stocks
belonging to the same economic sector is clearly seen.
Two stocks (lines) links when a horizontal line is drawn between
two vertical lines. The height of the horizontal line indicates the 
ultrametric distance at which the two stocks are connected.
The code of colors is the following: energy (blue), utilities (magenta),
financial (green), consumer/cyclical (brown), 
consumer/non-cyclical (yellow), conglomerates (orange), 
healthcare (gray), services (cyan), technology (red),
capital goods (indigo), 
basic materials (violet) and transportation (maroon).}
\label{fig1}
\end{figure}

\begin{figure}[t]
\epsfxsize=3in
\epsfbox{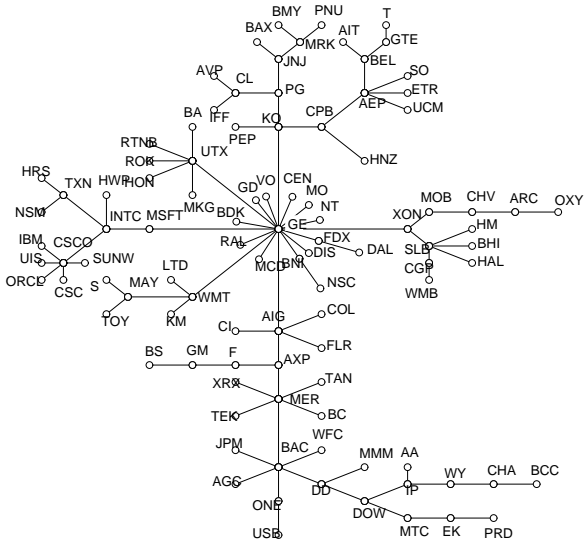}
\vspace{0.3cm}
\caption{Minimum spanning tree of the same set of stocks
and time horizon used in Fig. 1. Each circle 
represents a stock labeled by its tick symbol. The minimum 
spanning tree presents a large amount of stocks having a single link
and some stocks having several links. Some of these stocks act
as a ``center" of a local cluster. Examples are INTC and CSCO for 
technology stocks, AIG, BAC and MER for financial stocks
and AEP for utilities stocks. The stock GE (General Electric Co.) 
links a relatively large number of stocks belonging to various 
sectors.}
\label{fig2}
\end{figure} 

\begin{figure}[t]
\epsfxsize=3in
\epsfbox{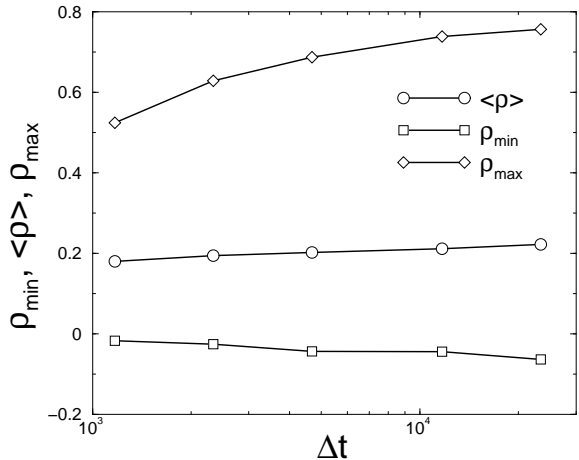}
\vspace{0.3cm}
\caption{Mean correlation coefficient $<\rho>$,
minimum and maximum value of $\rho$ as a function of $\Delta t$.
The values are obtained by averaging or searching 
over the $n(n-1)/2=4950$ off-diagonal elements of the
correlation coefficient matrix computed for the selected 
time horizon $\Delta t$. The degree of correlation 
or anti-correlation decreases by decreasing the time horizon.
The most prominent decrease is observed for the highest
levels of correlation ($\rho_{max}$).}
\label{fig3}
\end{figure}

\begin{figure}[t]
\epsfxsize=3in
\epsfbox{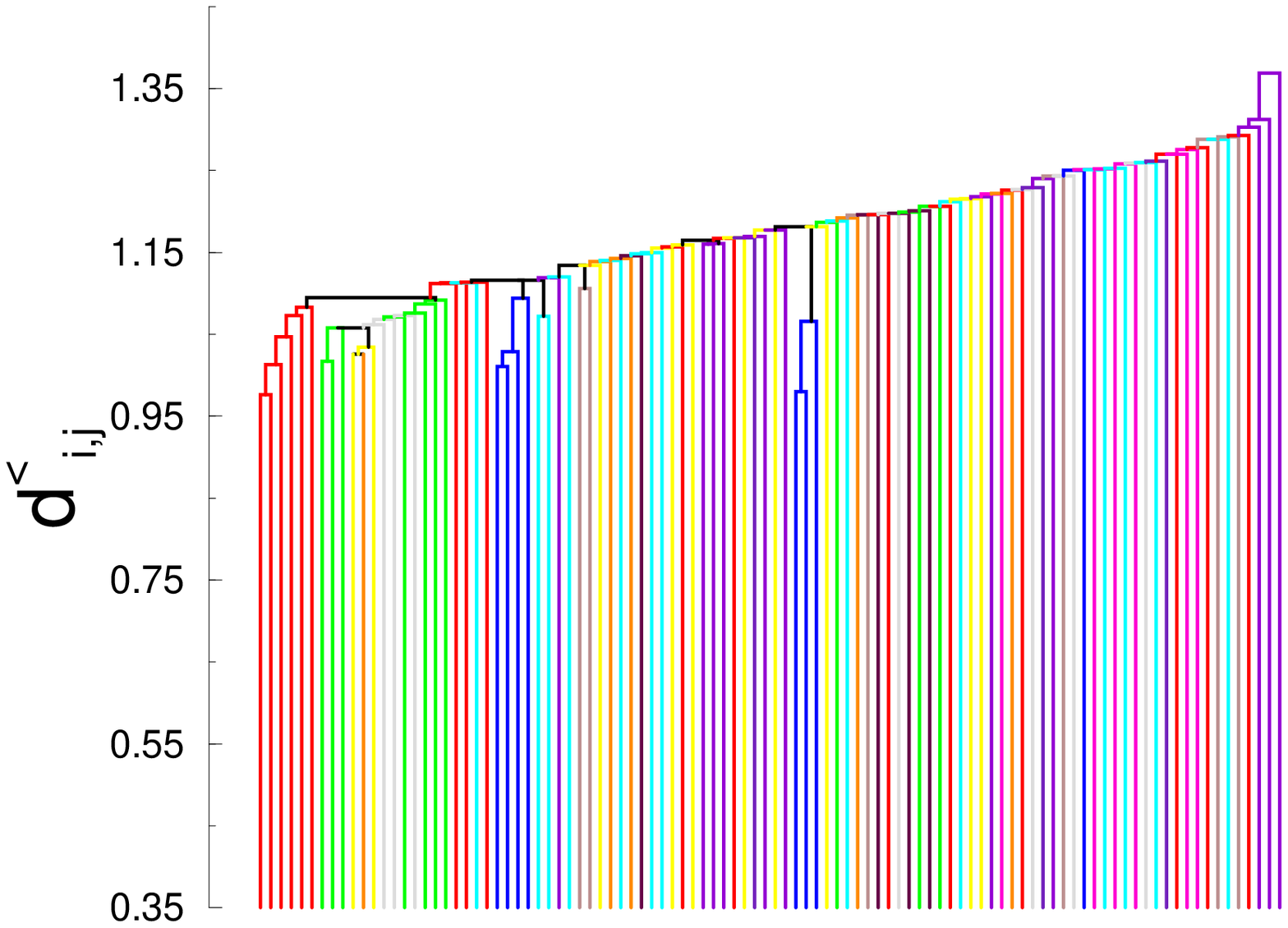}
\vspace{0.3cm}
\caption{Hierarchical tree of the set of 100 stocks traded in the
US equity markets obtained starting from the return time series 
computed with a
$\Delta t=19$ min and 30 s time horizon during the time period
Jan 1995-Dec 1998. Each stock is indicated by a vertical line and the
color of the line refers to its economic sector as defined by 
Forbes company. The presence of distinct clusters of stocks
belonging to the same economic sector is much less evident that
in Fig. 1. The code color is the same used for Fig. 1. The 
biggest clusters are the technology cluster
(red lines at the left side of the tree),
the financial cluster (green lines) and two energy 
clusters (blue lines).}
\label{fig4}
\end{figure} 

\begin{figure}[t]
\epsfxsize=3in
\epsfbox{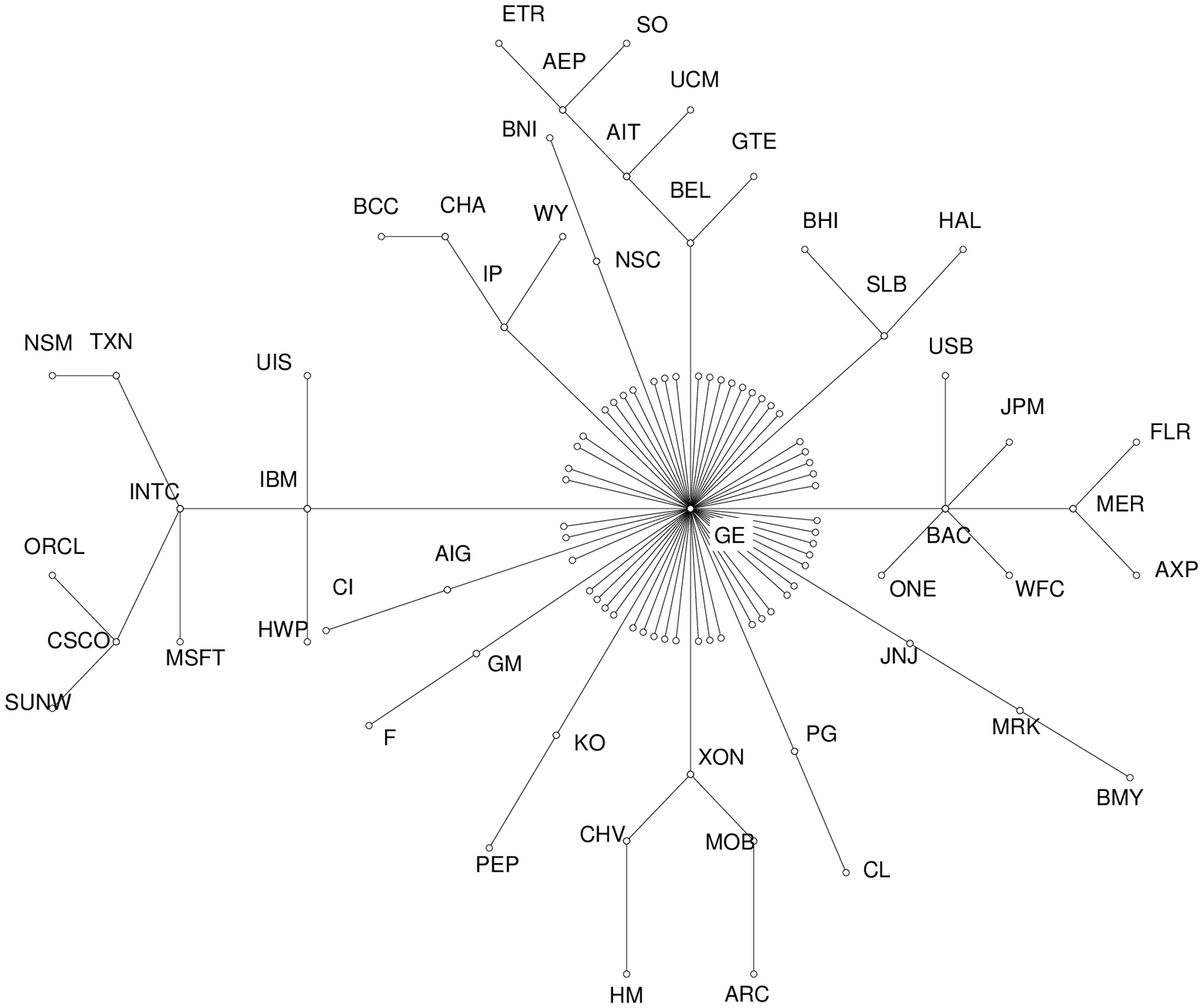}
\vspace{0.3cm}
\caption{Minimum spanning tree of the same set 
of stocks and time horizon used in Fig. 4. Each circle 
represents a stock labeled by its tick symbol. The minimum 
spanning tree presents a few cluster composed by companies
working in the same sector and by a large amount of stocks having 
a single link. Most of these stocks connect with a {\it reference}
stock that, in the present set, is GE.
The tick symbols of some stocks composing sector clusters are
shown in the figure. The technology cluster is
seen at the left side of the graph,
the financial cluster on the right side and two energy 
clusters are identified at the bottom and on the right-top
of the graph.
The tick symbols of the stocks directly 
connected with GE are: 
AA, AGC, AVP, BA, BAX, BC, BDK, BS, CEN, CGP, COL, CPB, CSC, DAL, DD, DIS, 
DOW, EK, FDX, GD, HNZ, HON, HRS, IFF, KM, LTD, 
MAY, MCD, MKG, MMM, MO, MTC, NT, OXY, PNU, PRD, RAL, 
ROK, RTNB, S, T, TAN, TEK, TOY, UTX, VO, WMB, WMT, XRX.}
\label{fig5}
\end{figure} 

\begin{figure}[t]
\epsfxsize=3in
\epsfbox{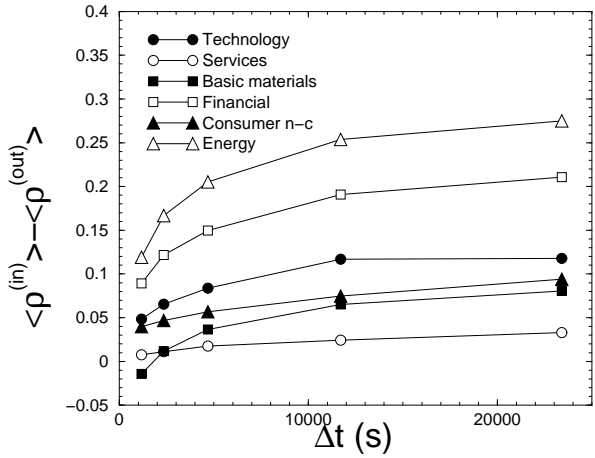}
\vspace{0.3cm}
\caption{Intrasector mean correlation coefficient $<\rho^{(in)}>$
minus the intersector mean correlation coefficient $<\rho^{(out)}>$
as a function of the time horizon for the largest groups of stocks
present in our investigated set. The ``Epps effect'' is 
more pronounced within each sector than outside it.
Groups are chosen by using
the classification of Forbes. The legend of used symbols is shown 
in the figure. For the number of stocks belonging to each group
see text.}
\label{fig6}
\end{figure} 

\begin{figure}[t]
\epsfxsize=3in
\epsfbox{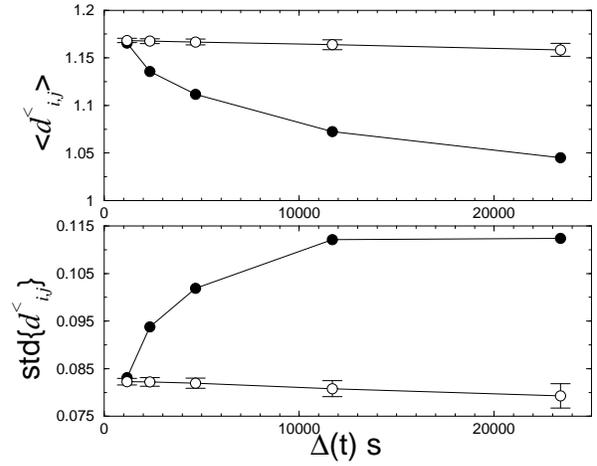}
\vspace{0.3cm}
\caption{Mean value (top panel) of the ultrametric distance 
between each pair of stocks connected in the MST and 
standard deviation (bottom panel) of the same quantity
for real data (black circles) and surrogate data (open circles)
generated starting 
from the correlation coefficient matrix computed for $\Delta t=19$ min
and 30 s. The error bars shown in the case of surrogate data are obtained
by determining the standard deviation of the investigated quantity
over 100 different realizations. By increasing the time horizon 
the mean value of $d^<_{i,j}$ decreases for real data whereas it remains
essentially constant for surrogate data. This implies that the 
degree of correlation of the detected hierarchical structure 
increases when $\Delta t$ increases only for real data.
By increasing $\Delta t$ 
the standard deviation of $d^<_{i,j}$ increases for real data 
whereas it slightly decreases for surrogate data. This 
behavior shows that the 
complexity of the clustering of the detected hierarchical structure 
increases when $\Delta t$ increases only for real data.}
\label{fig7}
\end{figure}

\begin{figure}[t]
\epsfxsize=3in
\epsfbox{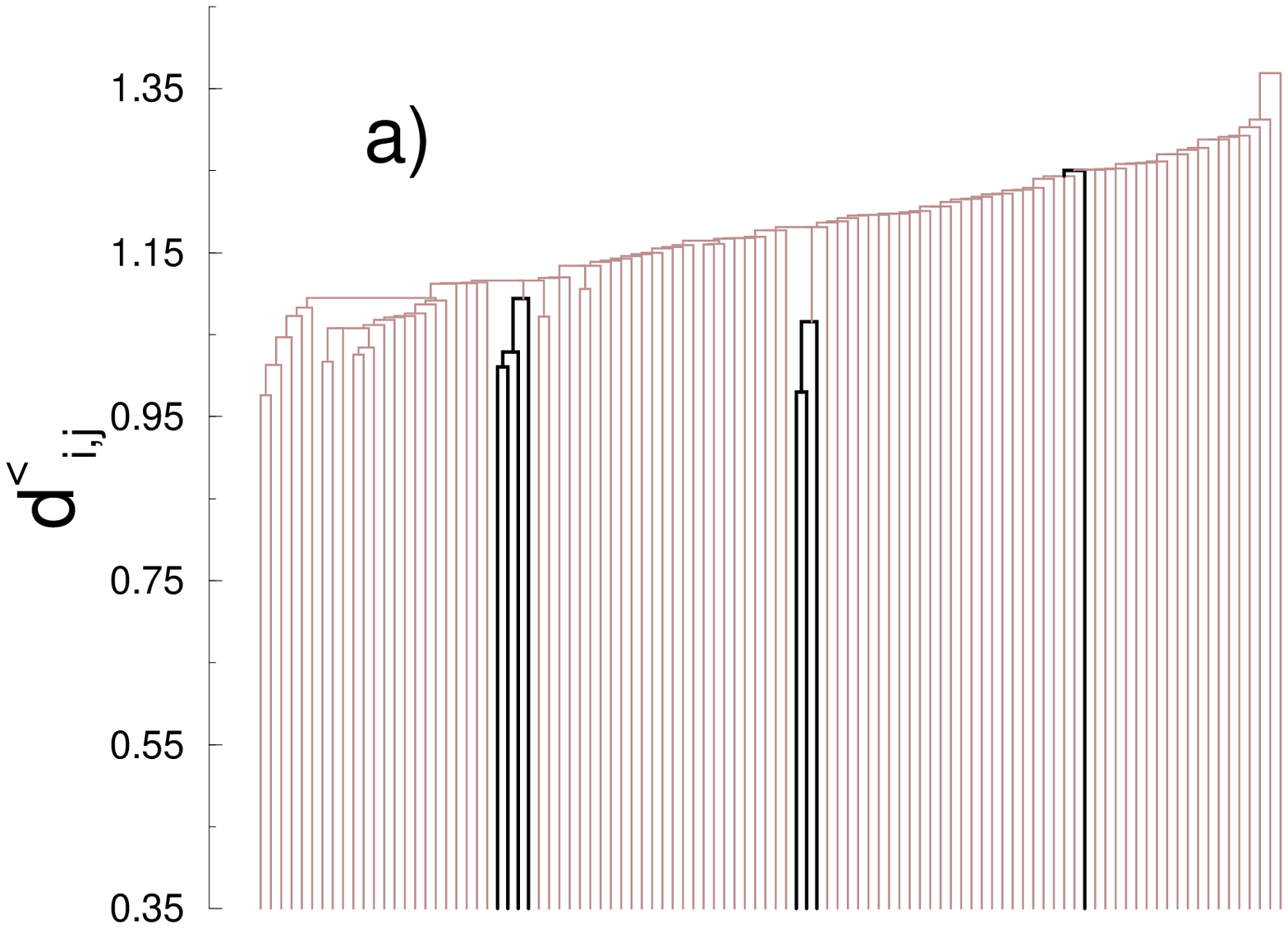}
\vspace{0.3cm}
\epsfxsize=3in
\epsfbox{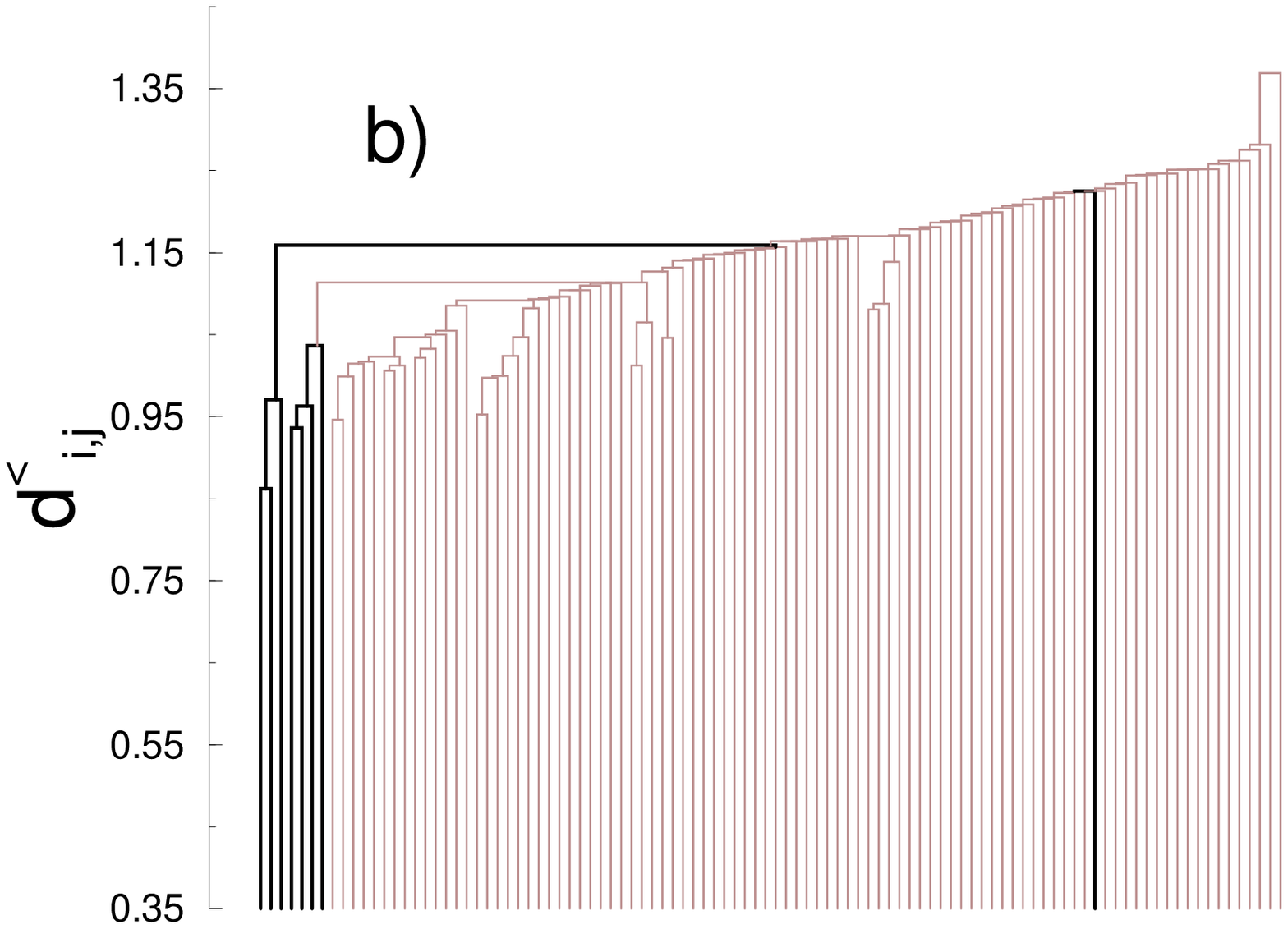}
\vspace{0.3cm}
\epsfxsize=3in
\epsfbox{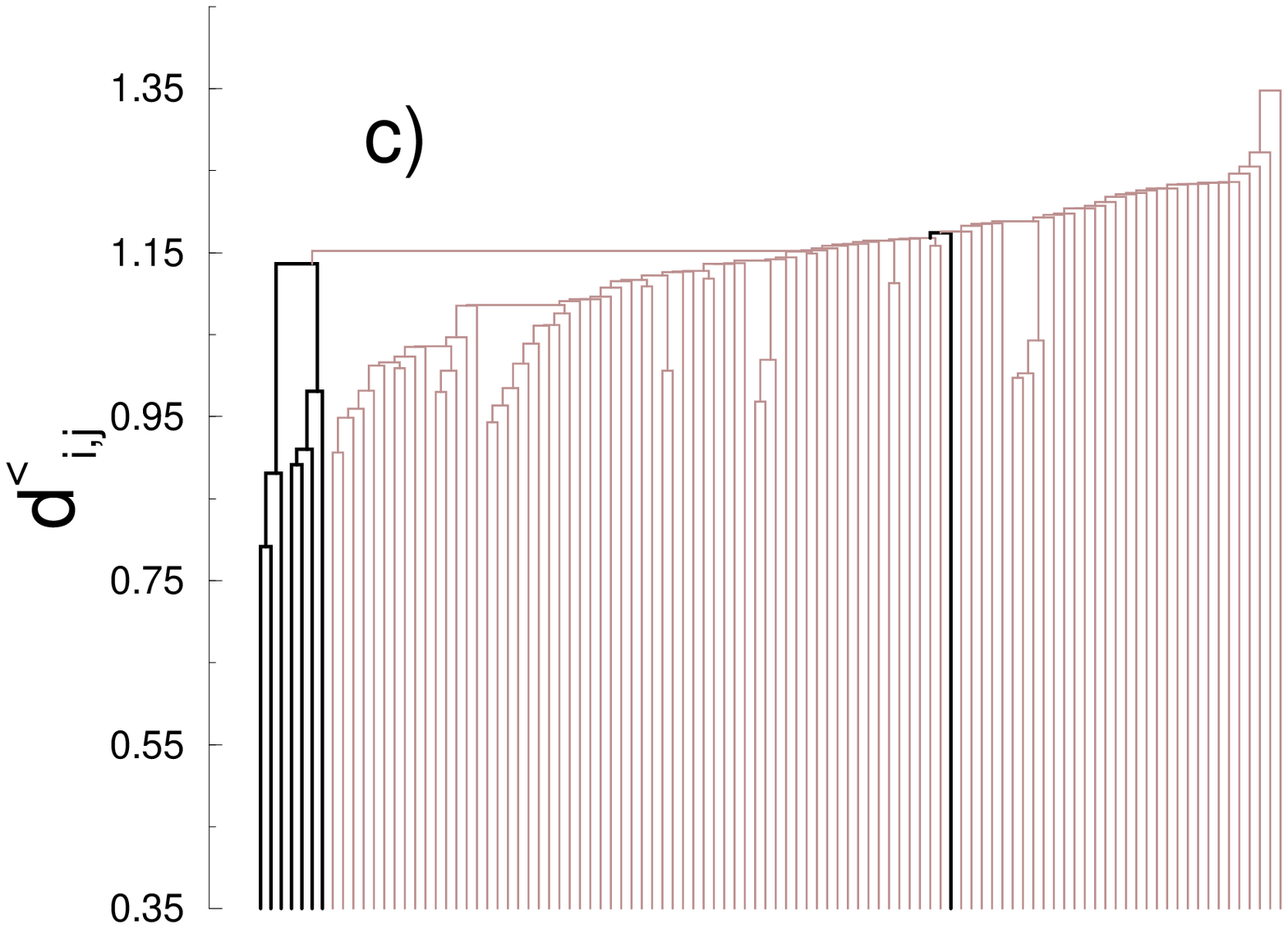}
\vspace{0.3cm}
\epsfxsize=3in
\epsfbox{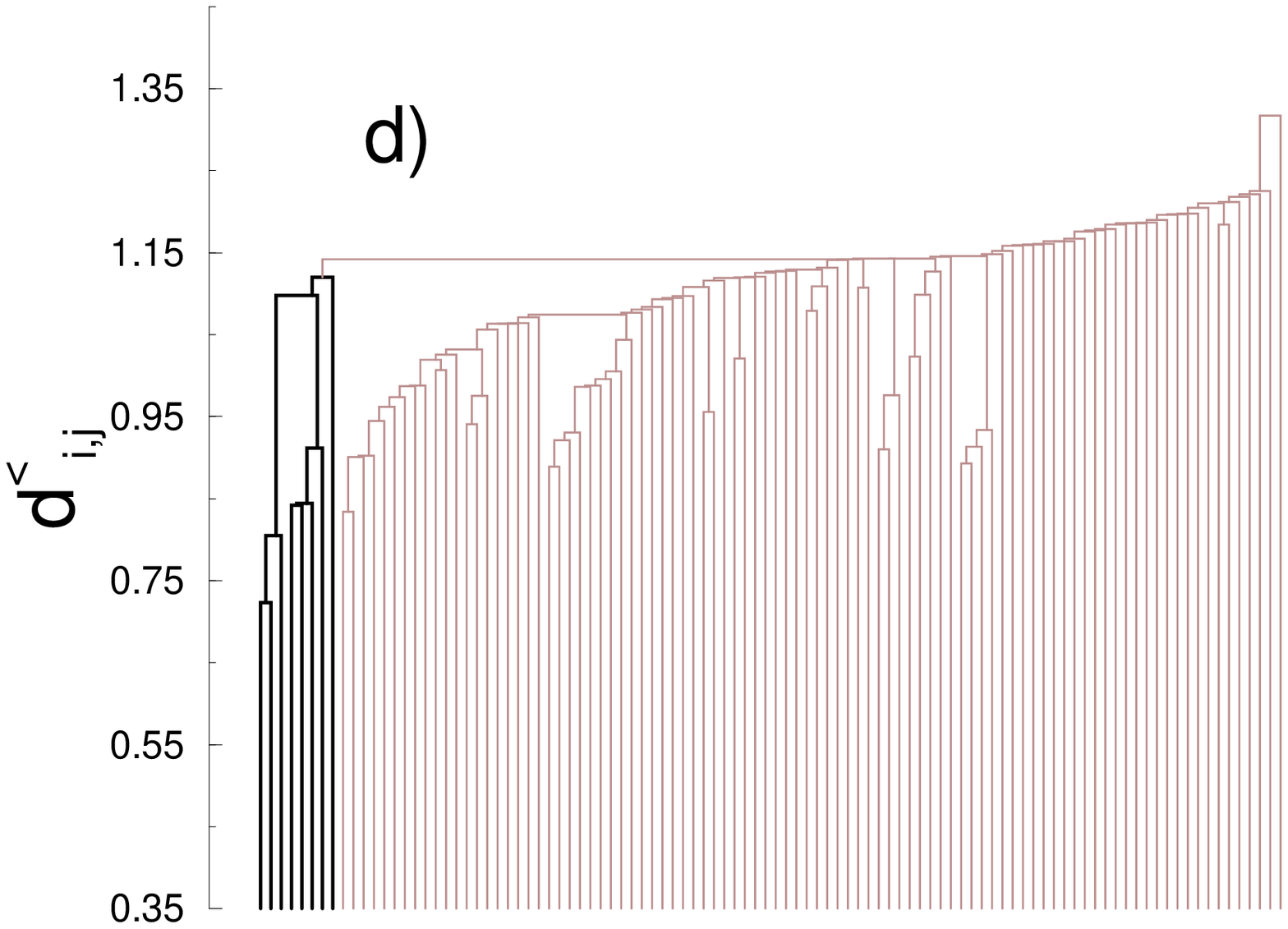}
\vspace{0.3cm}
\epsfxsize=3in
\epsfbox{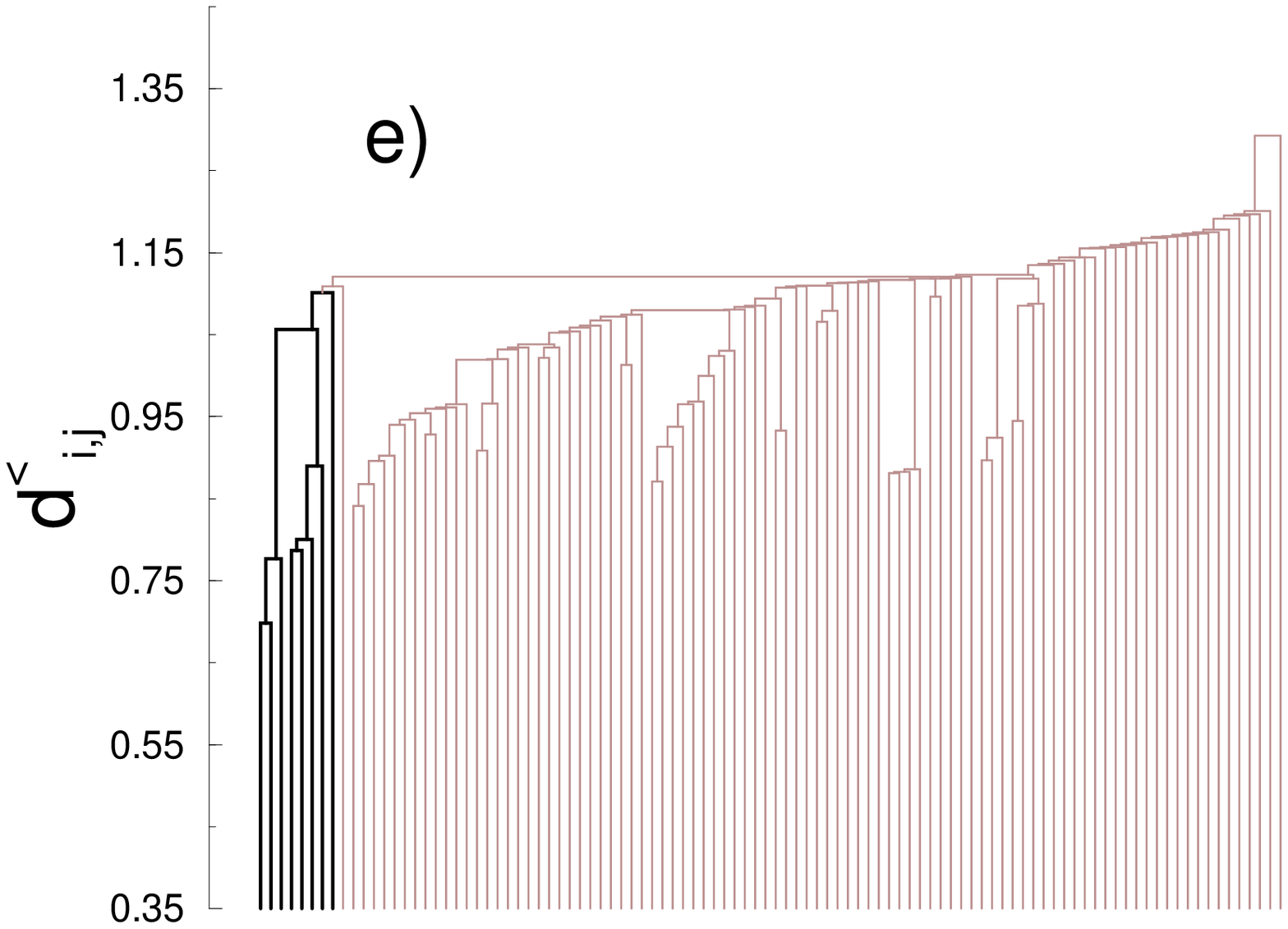}
\vspace{0.3cm}
\caption{Progressive structuring of the energy cluster in the 
hierarchical trees as a function of the time horizon.
The hierarchical trees are shown for (a) $\Delta t=d/20=19$ min and 30 s,
(b) $\Delta t=d/10=39$ min, (c) $\Delta t=d/5=1$ h and 18 min, 
(d) $\Delta t=d/2=3$ h and 15 min and (e) $\Delta t=d=6$ h and 30 min  
(one trading day time interval).
Black lines indicate 
stocks belonging to the energy sector whereas the remaining
lines are left gray.  
In Fig. 6a ($\Delta t=19$ min and 30 s) two distinct energy clusters are
observed. The first cluster is composed of XON, MOB, CHV and ARC, 
whereas the second comprises SLB,HAL and BHI. The remaining
energy stock OXY is disjointed from the two clusters and connected to 
the 'reference' stock GE. In Fig. 6b ($\Delta t=39$ min) almost
the same behavior as in (a) is observed. 
When $\Delta t=1$ h and 18 min (Fig. 6c), the two previous clusters 
merge into a single one which is distinct from all the others 
up to a ultrametric distance of $d^<=1.15$. 
In Fig.6d the OXY stock joins the energy cluster 
which maintains its internal
structure. In Fig. 6e we finally show the hierarchical 
tree obtained by using a 1 day time horizon ($\Delta t=6$ h and 30 min).
The energy cluster contains ,linked at a longer 
ultrametric distance, an additional stock (CGP) 
belonging to the natural gas utilities sub-sector.}
\label{fig8}
\end{figure} 

\begin{table}
\caption{Number of links in the minimum spanning tree as a 
function of the time horizon used to compute return time series. 
The stocks listed
are the stocks having more than one link when $\Delta t=6$ h and 30 min. 
Several stocks present a number of links that diminish when the
time horizon decreases from $\Delta t=6$ h and 30 min to 
$\Delta t=19$ min and 30 s. 
The behavior opposites to this general trend is the one
of GE. The number
of links of GE increases gradually from 17 to 
61 when $\Delta t$ decreases from 6 h and 30 min to 19 min and 30 s.}
\label{tab:1}       
\begin{tabular}{lccccc}
\hline\noalign{\smallskip}
tick symbol & d/20 & d/10 & d/5 & d/2 & d \\
\noalign{\smallskip}\hline\noalign{\smallskip}
GE	&	61	&	49	&	31	&	20	&	17	\\
BAC	&	6	&	7	&	8	&	7	&	6	\\
CSCO	&	3	&	3	&	4	&	5	&	6	\\
MER	&	3	&	5	&	6	&	6	&	6	\\
UTX	&	1	&	1	&	1	&	5	&	6	\\
AEP	&	3	&	4	&	4	&	4	&	5	\\
AIG	&	2	&	3	&	5	&	5	&	5	\\
SLB	&	3	&	3	&	3	&	3	&	5	\\
INTC	&	4	&	5	&	5	&	4	&	4	\\
KO	&	2	&	2	&	3	&	5	&	4	\\
WMT	&	1	&	2	&	3	&	3	&	4	\\
AXP	&	1	&	1	&	3	&	3	&	3	\\
BEL	&	3	&	4	&	5	&	4	&	3	\\
CL	&	1	&	1	&	1	&	3	&	3	\\
CPB	&	1	&	1	&	1	&	2	&	3	\\
DD	&	1	&	2	&	2	&	3	&	3	\\
DOW	&	1	&	1	&	2	&	3	&	3	\\
IP	&	3	&	3	&	3	&	3	&	3	\\
JNJ	&	2	&	2	&	3	&	3	&	3	\\
May	&	1	&	1	&	2	&	3	&	3	\\
MRK	&	2	&	2	&	3	&	3	&	3	\\
PG	&	2	&	4	&	4	&	2	&	3	\\
TXN	&	2	&	2	&	2	&	4	&	3	\\
XON	&	3	&	5	&	4	&	3	&	3	\\
ARC	&	1	&	1	&	3	&	2	&	2	\\
CGP	&	1	&	1	&	2	&	2	&	2	\\
CHA	&	2	&	2	&	2	&	2	&	2	\\
CHV	&	2	&	1	&	1	&	1	&	2	\\
EK	&	1	&	1	&	1	&	2	&	2	\\
F	&	1	&	2	&	2	&	2	&	2	\\
FDX	&	1	&	1	&	1	&	2	&	2	\\
GM	&	2	&	1	&	1	&	1	&	2	\\
GTE	&	1	&	1	&	1	&	1	&	2	\\
MOB	&	2	&	2	&	3	&	5	&	2	\\
MSFT	&	1	&	1	&	1	&	3	&	2	\\
MTC	&	1	&	1	&	1	&	2	&	2	\\
NSC	&	2	&	2	&	2	&	2	&	2	\\
ONE	&	1	&	1	&	2	&	1	&	2	\\
WY	&	1	&	1	&	1	&	2	&	2	\\
\noalign{\smallskip}\hline
\end{tabular}
\end{table}

\end{document}